# *Time-frequency analysis assisted reconstruction of ruthenium optical constants in the sub-EUV spectral range 8 nm – 23.75 nm*

Qais Saadeh[1], Philipp Naujok[2], Vicky Philipsen[3], Philipp Hönicke[1], Christian Laubis[1], Christian Buchholz[1], Anna Andrle[1], Christian Stadelhoff[1], Heiko Mentzel[1], Anja Schönstedt[1], Victor Soltwisch[1], and Frank Scholze[1]

[1]Physikalisch-Technische Bundesanstalt (PTB), Abbestraße 2-12, 10587 Berlin, Germany

[2]OptiX fab GmbH, Hans-Knöll-Str. 6, 07745 Jena, Germany

[3]IMEC, Kapeldreef 75, B-3001 Leuven, Belgium

Corresponding author: qais.saadeh@ptb.de

*Abstract*

**The optical constants of ruthenium in the spectral range 8 nm – 23.75 nm with their corresponding uncertainties are derived from the reflectance of a sputtered ruthenium thin film in the Extreme Ultraviolet (EUV) spectral range measured using monochromatized synchrotron radiation. This work emphasizes the correlation between structure modelling and the reconstructed optical parameters in a detailed inverse-problem optimization strategy. Complementary X-ray Reflectivity (XRR) measurements are coupled with Markov chain Monte Carlo (MCMC) based Bayesian inferences and quasi-model-independent methods to create a model factoring the sample's oxidation, contamination, and surface roughness. The sensitivity of the modelling scheme is tested and verified against contamination and oxidation. A notable approach mitigating the high dimensionality of the reconstruction problem is elaborated with the results of this work compared to two previously published datasets. The presented dataset is of high interest for the continuing development of Extreme Ultraviolet Lithography (EUVL) and EUV astronomy optical systems.**

1. Introduction

EUV Lithography (EUVL) systems are now used in high-volume manufacturing at the 7-nm-node. Its capabilities need to be further advanced and current challenges to be mitigated to support the future technology nodes. The characteristics of EUV radiation, with materials being highly absorbing requires the utilization of reflective optics and creates specific challenges for EUVL. To further improve the capabilities of EUVL and to examine the potential enhancement by materials optimization processes, numeric simulations are carried out [1]. The optical constants of the involved materials, $n(\lambda) = 1-\delta(\lambda) +i\beta(\lambda)$ are a critical input. $\delta$ and $\beta$ relate to the phase velocity and to the absorption of electromagnetic radiation with a wavelength $\lambda$, respectively. A hindrance for EUVL development is that the optical constants for many metallic materials in the EUV spectral range are not aptly determined. Inconsistencies are reported between different references for some materials [2,3]. In addition to the lack of calculated uncertainties for the optical constants of most materials, the accurate determination of the optical constants for materials becomes highly relevant for the development of next-generation EUVL optical components.

One technologically highly relevant and promising material for EUVL is ruthenium. It has been widely used in many critical optical components of EUVL systems [4]. Due to its high chemical stability and oxidation resistance, high reflectivity in the EUV spectral range and high etch selectivity to the materials of mask absorber layers (e.g.: TaN), Ru thin films are used as capping layers of the Mo/Si multi-layer mirrors of the masks, extending the lifetime of the deposited multi-layer mirror (MLM) reflector and protecting it during pattern repair and cleaning processes [4,5]. In addition to the outstanding performance of Ru capping layers in EUVL masks, Ru has been used and investigated to enhance the performance of other key EUVL components as it can be used for example in the coating of the inner reflective surface of the EUV grazing-incidence collectors and due to its high thermal emissivity Ru coated pellicles are reported to have better performance characteristics [4]. Recently, Ru and Ru-based alloys have been reported to be promising to mitigate the induced 3D mask imaging effects which is crucial for the development of next-generation EUVL technology. This could be achieved by using thinner EUV mask absorbers than the currently used tantalum-based absorbers [6].

The optical constants in the spectral range 8 nm – 23.75 nm (52.2 eV – 154.9 eV) are reconstructed from angle-dependent EUV reflectometry (EUVR) measurements conducted in the radiometry laboratory of the PTB at the electron storage ring BESSY II. The data was collected regarding a large angular range from near normal to grazing incidence, so the optimization yields results with high uniqueness in determining the optical constants of the thin film [7]. The EUVR method was selected since it allows non-destructive characterization of the samples with minimal preparation requirements, enabling the simultaneous reconstruction of $\delta$ and $\beta$. Another advantage of EUVR is that it allows the characterization of thin films optical characteristics with similar settings of their intended application. To aide in the reconstruction of the optical constants, complementary XRR data was used to construct a physical model of the sample structure in a quasi-model-independent formalism. Given the high sensitivity of XRR for the dimensional characteristics and interfacial imperfections of thin films, a reliable physical model of the sample can be further refined for the optical constants' reconstruction.

We put special emphasis on the determination of the parameter uncertainties using MCMC-based Bayesian inferences, with focus on the correlations between the reconstructed optical constants and the presumed sample's structure. Additionally, the credibility of the modelling scheme is substantiated by comparing the reconstrued geometrical and optical characteristics of the sample before and after a hydrogen radical cleaning process.

This dataset may also be useful for the materials dielectric spectroscopy community and for the development of EUV astronomy optics, since the investigated spectral range covers the emission lines of some ionized states of abundant elements that are considered in Solar weather research and of interest in planetary magnetospheres observations [8, 9].

2. Experimental setup

2.1 Sample preparation

The ruthenium thin film was deposited using DC magnetron sputtering onto a 300 mm Si wafer with a surface roughness of less than 0.1 nm root-mean-square (RMS). After evacuating the sputtering chamber ($p_0 < 10^{-7}$ mbar) argon with a high purity (6N) was used as a sputtering gas. The native oxide of the substrate was not removed before deposition of the ruthenium film. Afterwards, the coated wafer was cleaved into smaller square coupons with a side length of 25 mm. By manually analyzing the peaks of an XRR profile measured at Cu-K$\alpha$ regarding a witness sample, the XRR profile's critical angle was calculated and the approximated mass density from the electron density of the ruthenium film was found to be close to that of the bulk of 12.1 g/ cm$^3$ [10, 11].

2.2 EUV reflectivity measurements

The EUV reflectivity (EUVR) data was collected at the soft X-ray radiometry beamline (SX700) in the EUV radiometry laboratory of PTB at the electron storage ring facility BESSY II [12, 13]. The main characteristics of the SX700 beamline are shown in table1. A vacuum tank housing a reflectometer setup serves as an experimental endstation for the beamline to allow the necessary handling of the samples. The reflectometer's rotatable sample holder has a 6-axis goniometer, different orientations of the sample's surface can be irradiated while a wide angular range is covered between 1.5º and 90º relative to the surface's normal.

The reflectometer is fitted with GaAsP photodiodes on a movable arm synchronized with the goniometer enabling the specular reflectance scans with the two linear polarization settings S- and P- polarization. To reduce the potential contamination of the probed optical surfaces the reflectometer machineries are lubricant-free, and the vacuum tank is housed in a clean room environment.

Additionally, the SX700 reflectometer is equipped with a load-lock chamber to allow in vacuo transfer of samples to an auxiliary process chamber where hydrogen-radical cleaning is feasible [15]. In this chamber's setup, hydrogen radicals are produced by thermal dissociation of hydrogen molecules using a heated tungsten filament.

Table 1: Main parameters of the SX700 beamline at the standard settings [14].

| Parameter | Value |
| --- | --- |
| Wavelength Range | 0.7 nm to 24.8 nm |
| Beam Spot Size | 1 mm × 1 mm (variable) |
| Beam Divergence | 1.6 mrad × 0.4 mrad |
| Linear Polarization | 98.7% |
| Diffuse scattered light | 0.2% |
| Radiant power | 0.5 µW at 13.5 nm |

2.3 XRR Measurement

Before the cleaning procedure, XRR measurements on the same sample were performed at the plane grating monochromator (PGM) beamline of PTB [16], which provides soft X-ray radiation of high spectral purity in the photon energy range of 78 eV to 1860 eV. The employed experimental end station is an ultra-high vacuum chamber equipped with a multi-axis manipulator allowing for very precise sample alignment with respect to all degrees of freedom [17]. This endstation also enables XRR experiments in θ-2θ geometry in a broad angular range. The present sample was measured using an incident photon energy of 1750 eV in an incident angular range of 0° to 13°. Part of the obtained experimental data is shown in figure 2(a).

3. Bayesian framework for optical constants reconstruction

The reconstruction of optical constants from reflectance data is possible by the optimization of an inverse problem. Traditionally, this is achieved by iteratively minimizing an objective function that compares the measured dataset against a simulated one, based on a presupposed model. Where the parameters of the model are refined iteratively and are assumed to be descriptive of the probed sample once the stopping criterion of the optimization process is met. Preliminary results for the optical constants of ruthenium without uncertainties, based on such a classical optimization scheme have been presented before at a conference [18]. Ideally, this approach can be efficient when the initialized presupposed model aptly describes the sample's characteristics, but even then, it has drawbacks since the optimization space of reflectivity data is multi-extremal as there could be local minima so the definition of a stopping criterion is ambiguous [19]. Additionally, optimizing reflectivity data to reconstruct optical constants yields mere point estimates for each variable parameter in the problem. The uncertainties and the correlations of the reconstructed parameters are not retrieved in this case.

A framework mitigating the previously addressed complications and enabling an assertive approach for the reconstruction of optical constants, that is not heavily influenced by the initialized parameters, can be realized combining MCMC algorithms and Bayesian inferences. Earlier, the reconstruction of optical constants with their corresponding uncertainties from magnetron sputtered boron carbide thin films using MCMC, in the spectral range 40 nm – 80 nm has been conducted by treating each wavelength separately [20]. The attempt here regards a global simulation approach, presumably, this approach would yield higher accuracy, especially for the geometrical characteristics.

MCMC algorithms allow exploring the entire parameter sample space and Bayes' inferences aide in correlating the probabilities updated from the MCMC sampled space to informative reconstructed parameters. For this work, the MCMC sampler proposals are generated using the "Stretch Move" ensemble method proposed by Goodman and Weare [21,22].

From Bayes' theorem, the probability distribution of the parameters array p to represent the characteristics of the sample given the measured data (the posterior) is given by:

$$\Pr(\mathbf{p}|\mathbf{R}_{measured}) = \frac{\Pr(\mathbf{R}_{measured}|\mathbf{p}) \; \Pr(\mathbf{p})}{\Pr(\mathbf{R}_{measured})} \quad , (1)$$

The set of target parameters to be reconstructed from the measured reflectivity data are presented in the following array:

$$\mathbf{p} = [d_1, d_2, \ldots, d_n, s_1, s_2, \ldots, s_{n+1}, \delta_{\lambda,1}, \delta_{\lambda,2}, \ldots, \delta_{\lambda,n}, \beta_{\lambda,1}, \beta_{\lambda,2}, \ldots, \beta_{\lambda,n}, \ldots] \quad , (2)$$

Where $d_n$ and $s_n$ denote the nth layer thickness and the nth interfacial roughness, respectively, and $\Pr(\mathbf{R}_{measured}|\mathbf{p})$ is the probability distribution of obtaining the measured reflectivity data given the parameters array p (the likelihood). $\Pr(\mathbf{p})$ represents our (prior) knowledge about the sample and it is independent from any measurement. The evidence $\Pr(\mathbf{R}_{measured})$ is considered here as a normalization constant and hence (1) can be reformulated as:

$$\ln \Pr(\mathbf{p}|\mathbf{R}_{measured}) \propto \ln \Pr(\mathbf{R}_{measured}|\mathbf{p}) + \ln \Pr(\mathbf{p}) \quad , (3)$$

Assuming that the measurements errors have a Gaussian distribution and are not correlated, the likelihood (i.e.: logarithm of the likelihood) function evaluating the interrelations between the measured reflectivity and the simulated, supposing that the sample characteristics are represented by the parameters array p is given by [23]:

$$\ln \Pr(\mathbf{R}_{measured}|\mathbf{p}) = -\frac{1}{2} \sum_n \left[ \frac{(R_{simulated, \; n}(p) - R_{measured, \; n})^2}{\sigma_n^2} + \ln(2\pi\sigma_n^2) \right] \quad , (4)$$

Where $\sigma_n$ denotes the error in the nth measured data point and it is given here as a summation:

$$\sigma^2 = \sigma_{experimental}^2 + (a \cdot R_{simulated})^2 + b^2 \quad , (5)$$

$\sigma_{experimental}$ is the evaluated error of the measurement from the experimental setup, the two parameters a and b are supposed to compensate the effects of the model's deficiencies in describing the sample's real structure and the possible underestimation of the experimental error in our measurements, since it is very difficult to ascertain the effect of all the measurement's uncertainties on the measurands in such a complicated setting, these parameters are included in the parameters array p and to be sampled in the MCMC-based evaluation. In (5) we assume that the modelling error is normally distributed and has the same properties as the known measurement error.

To restrict the posterior from updating with probabilities from an overwhelmingly large range we define the prior as an uninformative conservative function for the set of parameters in p, so the likelihood search space is restricted to maintain higher efficiency for the MCMC algorithm:

$$\Pr(\mathbf{p}_n) = \begin{cases} 1, & \mathbf{p}_n \text{ lies in the expectation range} \\ 0, & \text{otherwise} \end{cases}, (6)$$

The desired result in the MCMC-based Bayesian inferences approach is the stationary distribution of the approximated posterior regarding all the parameters in the array p. A major limitation then is the required computational effort since it increases linearly with the number of variable parameters. The required number of MCMC proposals to reach the anticipated target distribution is usually considerable, especially with the large number of sampled parameters.

4. Hybrid model parametrization

The proper reconstruction of optical constants regarding thin films from reflectivity data is only attainable with the simultaneous reconstruction of the geometrical characteristics of the probed sample precisely. This poses a conspicuous obstacle since EUV radiation is highly attenuated by almost all materials. This limitation is exacerbated by the presence of inevitable contamination and possible surface oxidation concurrent with interfacial roughness and interdiffusion. To aide in the parameterization of the EUVR inverse problem, XRR measurements were opted for their non-destructiveness and high penetration depth enabling high sensitivity even for buried interfaces [24, 25].

A thin film sample is rarely accounted as single layer on a substrate system, due to oxidation, contamination, etc. Deducing information regarding a multi-layer from XRR data is not straightforward. Usually, extracting information is demonstrated with a time-consuming model-dependent approach utilizing trial and error optimization scheme. This becomes more laborious if none of the sample's characteristics is properly known a priori, or if the data is noisy which is often the case. Model-independent methods like Fourier analysis of XRR data can retrieve the density profile of a multi-layer system, aiding in setting a preliminary optimization model or to verify the results from one [11].

The Fourier transform (FT) of XRR profiles is a reliable method to obtain the number of layers and their corresponding thickness in a stratified system. The peaks of a FT spectrum of an XRR profile indicate the depths of the interfaces present in the sample, from which the layers thickness can be estimated. A limitation of FT is the loss of information regarding the temporal evolution in the case of a signal progressing in a time domain, which translates to the loss of information regarding the stratification order for XRR profiles with the intensity modulations progressing with increasing grazing incidence angle. Also, the FT resolution is vulnerable to the windowing procedure of the experimental data. Smigiel et al. [26] demonstrated overcoming the very information loss limitation using a time-frequency representation (TFR) that study the evolution of XRR profiles in the time and the frequency domains simultaneously, the short-time Fourier transform (STFT), enabling a quasi-model-independent approach to infer valuable information from phaseless XRR data. Similarly, Smigiel et al. [27] further elaborated the application of continuous wavelet transform (CWT) on XRR data where the properties of CWT produce a scalogram with higher quality and less noise enhancing the interpretability over the spectrogram obtained from STFT, enabling a better TFR, additionally allowing a model-independent estimation of the RMS roughness. Further details on the applications of time-frequency analysis on XRR data can be found in the references [26-28].

The approach utilized here to extract information from XRR data is hierarchical beginning with analysing a TFR based on the Wigner-Ville distribution (WVD) to infer the layers thicknesses and the stratification order (assuming a multi-layer system) [29]. This is followed by a fit using the global optimization metaheuristic algorithm Differential Evolution (DE) to arrive at refined values for the target parameters in particular the interfacial roughness [30]. Such an approach helps in deriving better and more reasonable results than solely using data fitting.

The WVD for a signal can be written as [31, 32]:

$$WVD(t, v) = \int s^*\left(t - \frac{\tau}{2}\right) s\left(t + \frac{\tau}{2}\right) e^{-2 i \tau v} d\tau \quad , (7)$$

Where $\tau$ is the lag variable, s(t) is the analytic signal derived from the real signal x(t) and $s^*(t)$ is the complex analytic, i.e.:

$$s(t) = x(t) + i\, H\{x(t)\} \quad , (8)$$

With H{x(t)} being the Hilbert transform of the real signal.

The WVD is used here due to its desirable properties as a bilinear TFR that decomposes a wave based on the density of its spectral content (energy), theoretically enabling high resolution readability when extracting features from the TFRs regarding non-stationary signals, that in some cases surpass the resolution obtained from CWT and STFT. The bilinearity of the WVD introduces cross-terms in the TFRs that usually appear between the signals' resolved features worsening the resolution, yet there are methods to eliminate those artefacts [33-35].

5. Data acquisition and analysis

5.1 EUVR Data

From the sample, two EUVR datasets (figure 1) were collected in the experimental endstation of the SX700 beamline by measuring the specular reflectance in the wavelength range 10 nm – 20 nm (in steps of 0.25 nm) for the angular range between 4.5° – 87° (grazing incidence, in steps of 1.5°).

Two datasets (EUVR datasets 1 and 2) were collected to test the sample's homogeneity. A total of 41 wavelength scans were taken in each dataset. The small difference between the two reflectivity maps regarding the different positions demonstrates the sample's high homogeneity (figure 1, c). The relative uncertainties of the measurements peak for steep incidence angle and high photon energy segment of the EUVR maps, i.e., for low reflectance values, where Kiessig fringes are observed (figure 1, b).

To enlarge the reconstruction range and to further examine the sensitivity of the modelling formalism against contamination and oxidation layers, an additional dataset (EUVR dataset 3) was collected from the sample after cleaning it with $H^*$. The presented dataset (figure 7) regards 64 wavelength scans in the spectral range 8 nm – 23.75 nm (in steps of 0.25 nm) for the angular range between 3° – 85.5° (in steps of 1.5°).

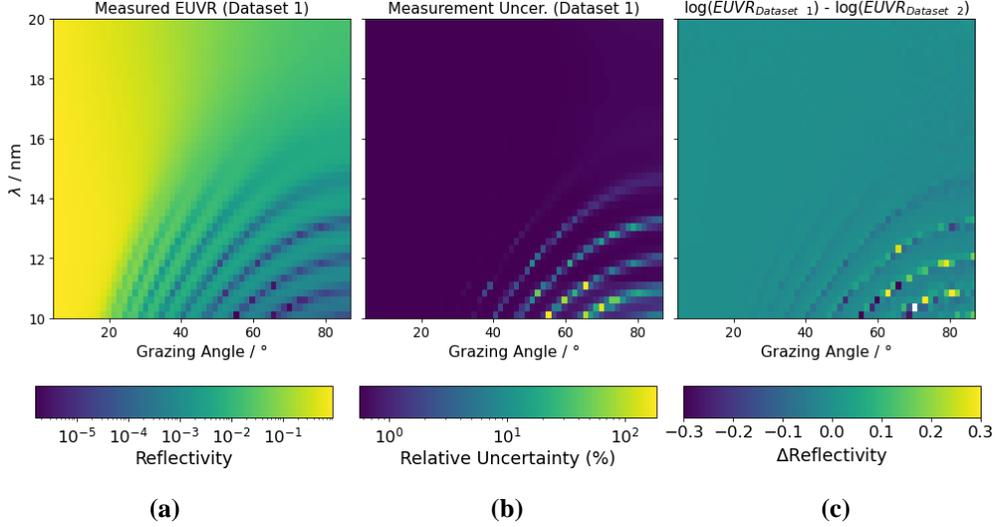

Fig. 1: Collected reflectance data from the SX700 beamline (a) measured EUVR from the center of the sample. (b) the relative measurements' error regarding the EUVR data shown in (a). (c) mapping of the difference between the two measured datasets, dataset 1 and 2.

5.2 Model parametrization

For a direct parametrization of the sample's structure the WVD (or its counterpart, the WV spectrum) of an XRR measurement at 1750 eV from the sample was calculated (figure 2) using multitaper spectral estimates followed with applying a reduced interference distribution (RID) kernel (see Krischer [36]), where the direct variable of the transform has been translated from the grazing angle to the *corrected* momentum transfer vector $Q_{z'}$ as given by [37-38]:

$$Q_{z'} = \frac{2}{\lambda}\sqrt{(cos\theta_{critical\ angle})^2 - (cos\theta_{grazing\ angle})^2} \quad , (9)$$

The conversion shown in (9) is to correct for the refraction. To obtain well-resolved features, the signal was renormalized prior to the transformation by taking $I(Q_{z'}) \cdot Q_{z'}^4$ [39].

Given the direct relation between the penetration depth of the X-ray incident beam in the probed sample and the grazing angle, one could argue that the features resolved at lower grazing angles (smaller scattering vectors) can be attributed to layers closer to the surface, while those features occurring at higher grazing angles are to indicate layers closer to the substrate [26]. Three features were resolved in the WVD (figure 2 (b), designated A, B and C). With the settings of the DC-magnetron sputtering tool adjusted at depositing a ruthenium film with a thickness of 50 nm, we could determine that feature A represents the targeted ruthenium deposition. With the sample being handled in the ambient conditions before the measurement, oxidation of the film and contamination from volatile molecules are expected. That is assumed to explain the presence of a top surface layer, feature B. Feature C could represent the substrate's native oxidation layer known previously from the manufacturer. Assuming feature C to represent the substrate's oxide, the discontinuity between the peaks of the two features B and C can be explained with the occurrence of the latter below another layer which is the ruthenium in this stratification.

For the interpretation of the stratification from the WV spectrum, we exclude a diffusion layer with the substrate since $SiO_2$ has a lower enthalpy of formation than ruthenium silicides, effectively enabling a diffusion barrier [40].

The gain from using time-frequency analysis here is the direct verification of the presence of a surface layer on top of the ruthenium film, this verification is critical since the reconstruction of optical constants is sensitive to the presence of such surface layers yielding different results depending on the assumed model as it will be demonstrated afterwards.

The surface layer resolved from the XRR data developed in an uncontrolled environment (ambient conditions). It is expected that this surface layer is comprised of the ruthenium film's oxide and additional physisorbed volatile molecules from the ambience. Surface ultra-thin layers could have a gradient owing to factors like the moisture and air, and it is assumed here to be the case [41]. The former assumption will be corroborated with the outcomes of the EUVR datasets. Optimally, modelling such a surface layer requires a transition-layer approach where a stack of transition layers whose density profile trails a smooth transition gradient that represents the examined layer [42]. This approach can model any roughness distribution, but it is very computationally intensive. The alternative here is to *effectively* model the surface layer as two separate layers, a carbonaceous contamination layer (with reduced density) above and oxide layer. This approximation is deemed necessary to reduce computational intensiveness since the EUVR data will be further treated regarding a four-layer model.

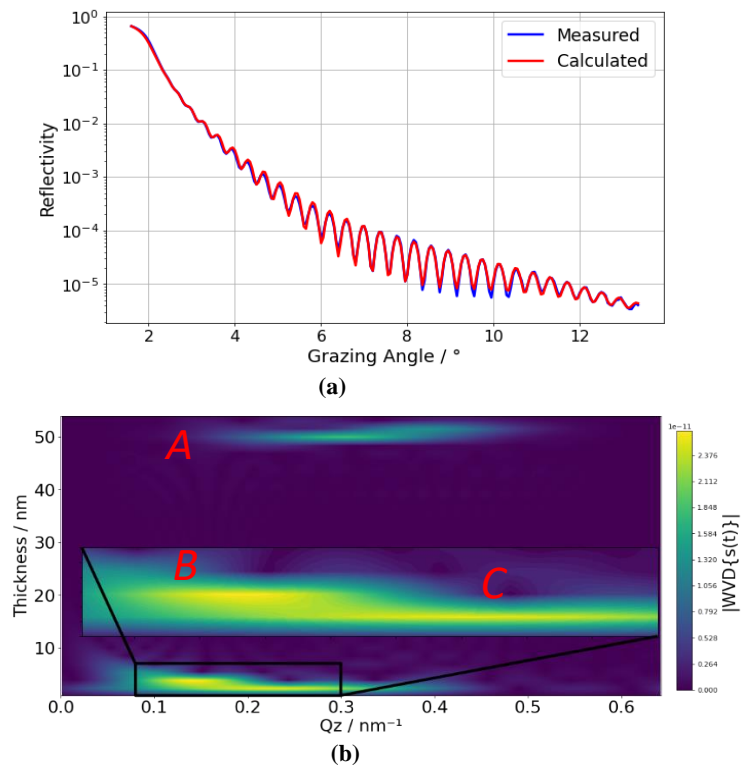

Fig. 2: The quasi-model-independent hierarchal approach to study the XRR profile of the sample. (a) The measured XRR profile at 1750 eV and its fitting upon calculating the WVD. (b) The absolute value of the WV spectrum for the of the XRR data. The red letters designate the apparent features. The magnified sub-plot emphasizes the discontinuity between two of the resolved features B and C.

A preliminary four-layer model (on a substrate) describing the sample structure can be presupposed, ascendingly, the substrate native oxide, ruthenium deposition, oxidation layer and a carbonaceous contamination. To model the interfacial imperfections the XRR data was optimized using the algorithm DE utilizing a logarithmic objective function. The forward calculation in the optimization process was simulated using Parratt's reflectivity formalism for layered systems with Névot-Croce damping factors introduced to compensate the effect of interfacial imperfections [43, 44]. The optical parameters of ruthenium and its oxide were left variables in the optimization given the anticipated reliable high sensitivity from a 50 nm layer for the former and to compensate for the uncertainties in the estimated parameters of the latter, since the exact oxide's stoichiometry ($RuO_x$) and its density are undetermined while the optical parameters of other materials were fixed from the Center for X-Ray Optics (CXRO) database [45].

5.3 Curse of dimensionality mitigation

The optimization of the discussed four-layer on a substrate model regarding the experimental settings of the two measurements corresponding to the spectral range 10 nm – 20 nm requires 424 parameters. With the parameters representing the optical constants, the structural characteristics, and other parameters relevant for the calculation (table 2). An even larger number for the optimization of dataset 3. Ideally, all parameters must be variables in the MCMC-Bayesian framework to reveal the correlations and to examine the propagation of uncertainties. With such a tremendous number of variables, applying MCMC-Bayesian inferences framework is impractical here as it requires an enormous computational effort for the chains to reach the equilibrium state for all parameters [46]. Since the aim here is the reconstruction of the optical constants of pure ruthenium and given the increased uncertainty in the reconstructed optical constants from ultra-thin films [47]. The optical constants of the ultra-thin layers and the substrate are to be substituted from previously published datasets (table 2 *merely* details for the case of datasets 1 and 2) [45, 48,49].

Table 2: The construction of the required parameters for the calculations relevant to the optimization of the EUVR datasets 1 and 2. V and C denote Variable and Constant, respectively. Array p includes the variable (91) parameters.

| Aspect | | Num. | Status | Note |
|---|---|---|---|---|
| Sample's structure / parameter | Layers thickness | 4 | V | Initialized from the WVD/ XRR optimization |
| | Interfacial roughness | 5 | C | From the XRR optimization |
| | $RuO_2$ density | 1 | V | Initialized from ref. [39] |
| | Ru density | 1 | C | *Fixed as 12.1 g/cm³ ,See section 2.1* |
| Optical constants and ASFs / material | Carbon | 82 | C | From ref. [43] |
| | $RuO_2$ | 82 | *See note* | Oxygen ASFs from ref. [43], Ru ASFs are variables |
| | Ru | 82 | V | ASFs Initialized From ref. [45] |
| | $SiO_2$ | 82 | C | From ref. [48] |
| | Si | 82 | C | From ref. [45] |
| Experimental settings (polarization and angular off-set) | | 2 | V | Initialized from ref. [12,13] |
| Uncertainty model's parameters | | 2 | V | Further details in ref. [22] |

Moreover, the optical constants of ruthenium oxide are to be *partially* fixed using the atomic scattering factors (ASF) representation since the complex index of refraction for a compound with N different atoms can be expressed as [45]:

$$n(\lambda) = 1 - \delta(\lambda) + i\beta(\lambda) = 1 - \frac{r_e}{2\pi} \lambda^2 \sum_{j=1}^{N} n_j \, f(\lambda)_j \qquad , (10)$$

Where $r_e$ is the classical electron radius, $\lambda$ is the wavelength, n is the number of atoms of type j per unit volume. f is the complex ASF for an atom with a real part related to the dispersion and an imaginary part related to the absorption:

$$f(\lambda) = f_1(\lambda) - if_2(\lambda) \qquad , (11)$$

With an assumed oxide stoichiometry of $RuO_2$ but variable density, the ASFs of ruthenium are now part of array p and the ASFs of oxygen are fixed as taken from the CXRO database. To further exclude additional variables in the forward calculation, the roughness parameters of the stratification are to be replaced with the RMS values obtained from the XRR fitting, considering that XRR measurements are better suited for indicating such values than EUVR measurements given their low penetration depth to probe the buried interfaces under the ruthenium deposition.

5.4 MCMC-Bayesian inferences from datasets 1 and 2

With 91 parameters left in array p 2.5e5 forward calculations based on Parratt's formalism coupled with Névot-Croce factors were iterated. The stationary target distribution of the MCMC chains was attained after ca. 3.6e4 iterations considering all the parameters. The calculations were done using emcee, a Python implementation of Goodman & Weare's Affine Invariant MCMC Ensemble sampler to generate the proposals regarding 500 *walkers* for each parameter [50].

The first 4e4 samples of the MCMC chains were discarded for the subsequent analysis (burn-in), since most of them have a low probability density to describe target parameters. To reduce autocorrelations and to tackle computer memory limitations, only every 15$^{th}$ step was saved from the MCMC chains (thinning of 15, figure 3(a)).

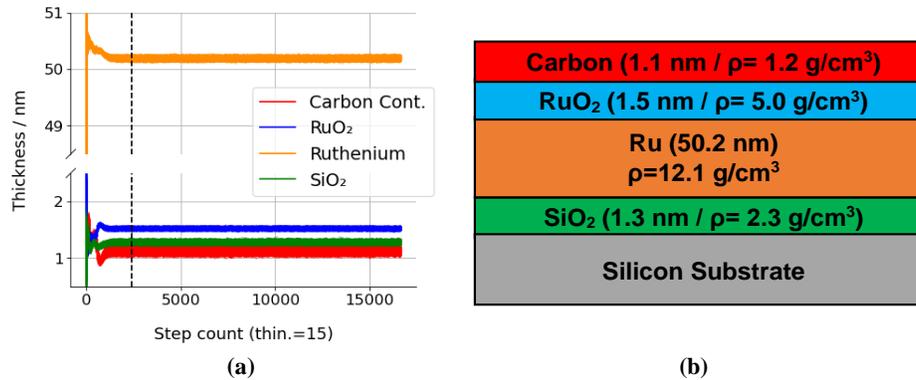

Figure 3: MCMC sampling (results) relevant for the geometrical characteristics. (a) Trace plots of the thickness parameters. The black dashed line marks the end of the burn-in period. (b) A sketch showing the reconstructed assumed stratification of the ruthenium sample with the mean values of the relevant chains. The ruthenium density is fixed to the nominal value. The dimensions are not scaled.

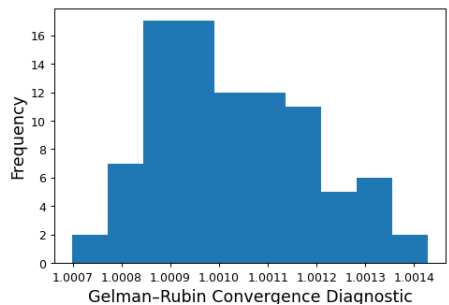

Figure 4: Gelman-Rubin criterion calculated for the sampled chains and projected as a binned histogram.

Another problem with applying MCMC calculations is the inability to clearly detect convergence. There is no unanimous agreement in the statistics community when it comes to the stopping criterion of an MCMC simulation, rather, empirical convergence diagnostics are usually used to check the possible convergence of the chains. Since a single diagnostic might fail, three diagnostics were checked for the convergence in this study: Gelman-Rubin criterion, autocorrelation analysis and Geweke-plots. All of which gave a positive *indication* for a potential convergence (figure 4) [51].

The calculated Gelman-Rubin factors for all the chains are below 1.1 which is the recommended cutoff value to stop MCMC sampling (figure 4). Geweke Z-scores were calculated with all found to be in the interval (-2,2). Another very important check is to examine the serial correlations of the samples given a chain, the autocorrelation function of the samples showed decreasing behavior and was oscillating around zero right before terminating the algorithm [51].

To emphasize the significance of the model used here, an additional MCMC simulation was applied on an oversimplified single-layer (on a substrate) model regarding EUVR dataset 2. The reconstructed optical constants from each model will be compared graphically (figure 8), albeit the discussion will merely focus on the four-layer model.

5.5 MCMC-Bayesian inferences from dataset 3

With 137 parameters left variables 1.2e5 forward calculations were iterated. The number of variables here is larger since dataset 3 has a larger spectral coverage than datasets 1 and 2. The stationary target distribution was attained after ca. 4.4e4 steps. The geometrical model retrieved from the modes of the chains (excluding the burn-in) is akin to that shown in figure 3(b) except for the thickness of the carbon contamination layer and the density of the oxide layer (figure 5). This verifies the sensitivity of the modelling scheme to surface layers and shows the anticipated slight increase in the base film's thickness upon the reduction of its oxide due to the cleaning process with hydrogen [52].

Since the XRR measurements were taken before cleaning the sample, the parameterization of the RMS values of the roughness parameters considering the two uppermost interfaces in the optimization of EUVR dataset 3 was slightly different than that of datasets 1 and 2. The roughness parameters of the two uppermost interfaces were scaled with the thickness of contamination layer by a reduction factor relating the thickness of the carbon layer obtained from dataset 1 (figure 3 (b)) to the thickness obtained from dataset 3 (the *online* chain's mode value during sampling). This approximation compensates the anticipated significant deformation of surface layers upon cleaning and is necessary since otherwise the roughness parameters would be on the same order of magnitude as the contamination layer.

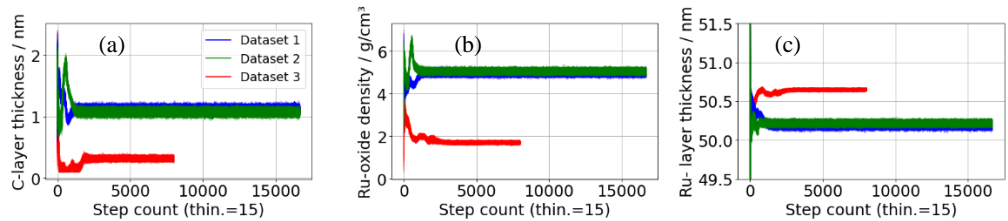

Figure 5: Trace plots of the parameters where significant difference is observed upon H$^*$-cleaning. (a) The thickness of the (assumed) carbon contamination layer decreases from ca. 1.1 nm to 0.3 nm. (b) The oxide's density drops by ca. 68% as retrieved from the trace plots relevant to dataset 3. (c) The observed increase in the thickness of the base film.

6. Reconstructed optical constants discussion

For the two datasets 1 and 2, the first 38.5 thousand iterations (burn-in period) were discarded, and the chains were integrated where each yielded Gaussian probability distributions with numerous parameters that exhibit correlations. For the reconstructed ASFs of ruthenium, the projected uncertainty defined as the confidence interval with 3-σ of the imaginary parts is generally higher than that of the real parts. Figure 6 shows the reconstructed ASFs of ruthenium regarding a wavelength of 13.5 nm, with their covariances and correlations to selected parameters.

The measurements' experimental uncertainty includes the uncertainty of (i) intensity measurement, (ii) stability of the synchrotron radiation and the beamline, (iii) and the diode noise. The MCMC-Bayesian framework has the polarization as a variable, to account for minimal unpolarized fraction of the radiation. The uncertainty induced from the alignment procedure was also compensated in the simulation by introducing a single angular off-set for each of the two datasets. Moreover, the effect of the beam's divergence was simulated on the EUVR data at a wavelength of 8.0 nm, and the difference was within the experimental uncertainty for the experimental data presented here. Merely the shortest wavelength was selected for the simulation of divergence effects, because Kiessig fringes are most susceptible to blurring and those fringes are the most prominent in our data at the wavelength of 8.0 nm [53]. Since the reflectance measurements are normalized with the detected synchrotron beam coming from the reflectometer's exit slit at each wavelength, the uncertainties due to calibration of the monochromator and the detector efficiency are assumed to be negligible. The two variables left in the weighted uncertainty (equation 5) are expected to compensate the effects of additional uncertainties on the reconstruction, such as the uncertainty induced from fixing the roughness parameters values. With a similar regard for the optical constants of the ultra-thin layers in the stratification, many parameters correlations with the projected posterior distributions of the reconstrued ASFs are lost. But this uncertainty modelling is more effective than calculating all the potential error sources [54]. Assuming that all of which can be identified in such a sophisticated experimental setting.

The reflectivity maps simulated using the modes of the posterior probability distributions show a remarkable resemblance with the measured datasets. For example, the goodness-of-fit is demonstrated for dataset 3 in figure 7.

The calculated optical constants from the reconstructed ASFs (equation 10) were compared with the optical constants retrieved from listed ASFs from two known published datasets (figure 8) [43,57]. The reconstructed optical constants from the three datasets are almost matching, although the reconstruction from each dataset was conducted separately. Generally, our data agree with those retrieved from the CXRO but showing deviations above 15 nm and both show disagreement with the optical constants retrieved from Chantler's database.

Also, the inferences from an oversimplified (single layer) model leads to significant variations above 15 nm (approximately). MCMC chains of parameters relevant to the experimental settings like the polarization were compared regarding the two models. The mode value in the case of the single-layer model is ca. 8.6% (linear polarization) where it doesn't exceed 2.7% for the case of a four-layer model, considering the 3 collected datasets. Presumably, this can be considered as another verification for accuracy of our model since the latter model yields more realistic results known of our experimental settings (table 1).

The prior range for the case of dataset 3 has been reduced, since the relatively large range used for the case of the two datasets 1 and 2 was more than sufficient. This decrease also speeds-up the calculations (green filled region in figure 8).

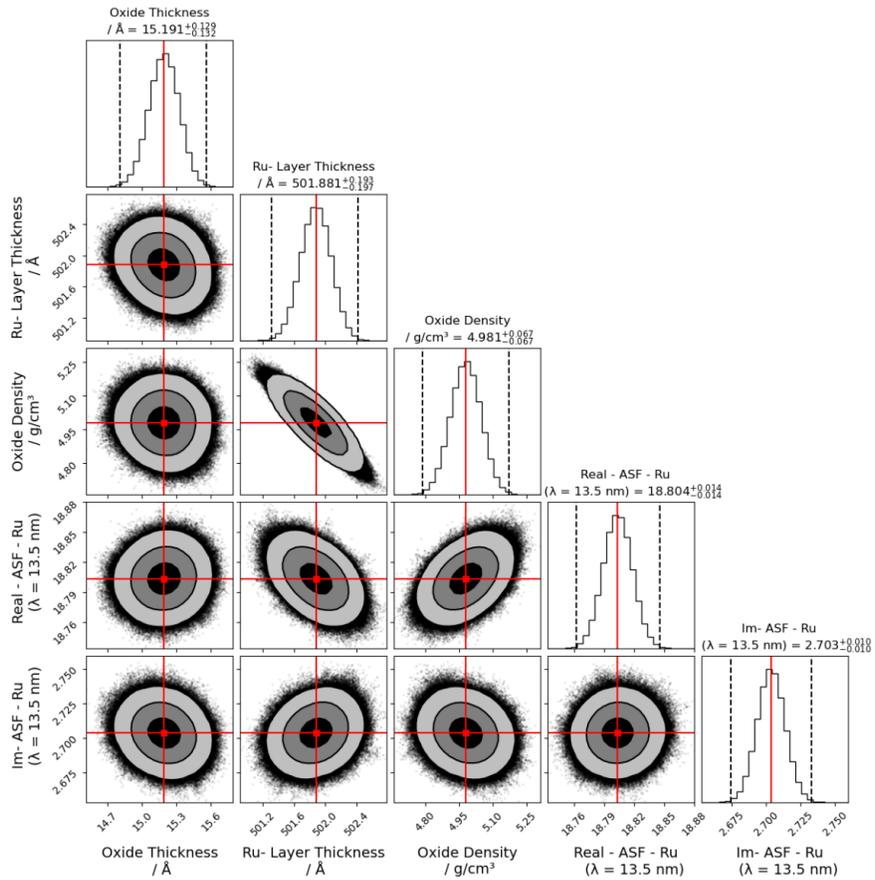

Figure 6: Corner plot of the posterior distributions from five selected sampled chains retrieved from dataset 1. The one- and two-dimensional projections of the sample chains are plotted to reveal covariances and correlations [56, 57].

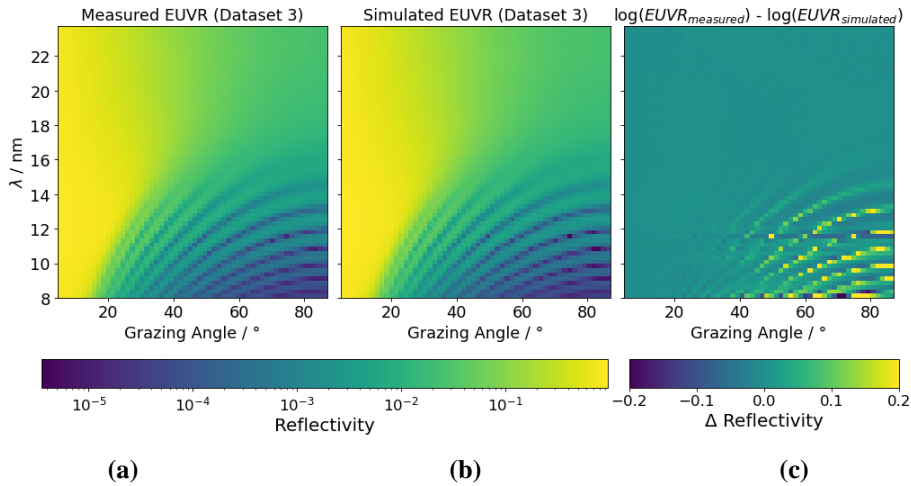

Fig. 7: Dataset 3, collected reflectance data after cleaning the sample and its simulation with the modes of the relevant MCMC chains. (a) collected EUVR data (b) reconstructed EUVR map (c) mapping of the residual between the measured and the simulated EUVR.

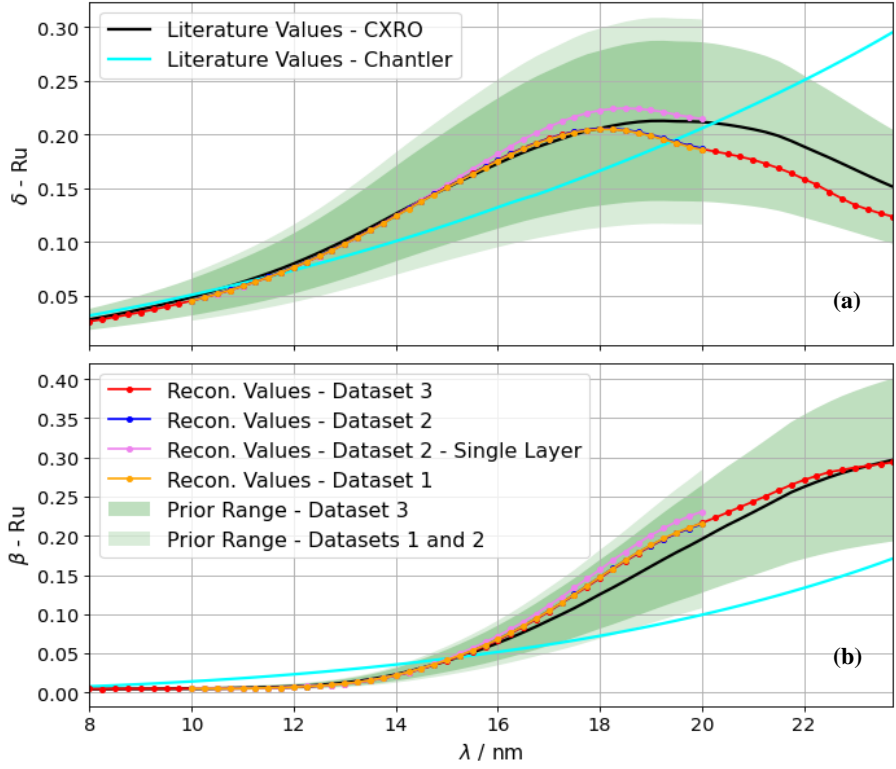

Figure 8: The plots represent the calculated optical constants from the reconstructed ASFs assuming a bulk's density with graphical comparison with the optical constants retrieved from listed ASFs from two previously published datasets. The green filled region is the expectation range of the optical constants which was initialized from the CXRO database (a) calculated dispersive (real) part (b) calculated absorptive (imaginary) part of the complex refractive index.

The significant drop in both the contamination layer's thickness and the oxide's density upon H* cleaning demonstrates the sensitivity of the adopted modelling scheme to surface contamination and oxidation. It also further supports the parametrized model from the WVD-based formalism.

The relative uncertainty in the reconstructed ASFs varies with the wavelengths, with small fluctuations regarding the AFS corresponding for $\lambda < 13$ nm. Those fluctuations can be explained with the presence of the highest measurement uncertainties in the spectral range where Kiessig fringes are observed (see figure 1, c). In addition to the uncertainty retrieved from the MCMC-Bayesian framework, we estimate 2% *overall* increase for the projected uncertainties of the reconstructed ASFs prior to calculating the estimated uncertainties of optical constants. This compensates for the uncertainty in the estimated nominal density of ruthenium (table 3), where the density of the ruthenium film was assumed to be that of the bulk density (12.1 g/ cm$^3$), which is the targeted density of the film from the magnetron sputtering deposition in our case [10].

7. Conclusions and potential outcomes

From reflectivity data collected in a synchrotron facility, we have presented the reconstruction of optical constants in the EUV sub-spectral range from 8 nm to 23.75 nm, using MCMC-Bayesian inferences framework with addressing the interval of confidence for each reconstructed parameter. The overall results were compared with their corresponding in two published datasets and our results show resemblance to the CXRO data with deviations above approximately 15 nm.

We successfully applied the WVD based TFR on experimental XRR data. The significance of resolving surface layers directly from the experimental data without prior knowledge is emphasized in the model selection [25]. The outcome of this work encourages further applications of time-frequency analysis given its ability to resolve ultra-thin layers directly, especially surface layers. This is of crucial value since the optical characteristics of thin films in the EUV spectral range are highly sensitive to the presence of such surface layers that are not easily resolved with a trial-and-error analysis of the XRR data. Summarizing, time-frequency analysis is shown to be a valuable metrology tool for quasi-model independent investigation of thin film stacks.

The reliability of applying MCMC-Bayesian inferences was tested and verified, with emphasis on the reconstruction's sensitivity for surface layers, where three different datasets yield very close values for the reconstructed ASFs, but significant differences are observed in the surface layers characteristics upon cleaning the sample.

Finally, we applied a new comprehensive approach to the determination of optical constants from thin film reflectance data. Ruthenium chosen here as an example is particularly interesting for EUVL as it is used in many optical components there. The presented analysis scheme is based on open-source and highly accessible software packages, promoting inverse-problem solving strategies of EUVR data. Based on the presented scheme, this work will be followed by publications utilizing updated strategies developed for the reconstruction of optical constants of other materials, where the presented strategy circumvents the requirements of retrieving the optical constants using Kramers-Kronig relations, which is often hindered by missing optical constants in parts of the electromagnetic spectrum.

*Funding & Acknowledgements*

This project has received funding from the Electronic Component Systems for European Leadership Joint Undertaking under grant agreement No 783247 – TAPES3. This Joint Undertaking receives support from the European Union's Horizon 2020 research and innovation program and from Netherlands, France, Belgium, Germany, Czech Republic, Austria, Hungary and Israel.

*Appendix*

The reconstructed optical constants obtained from dataset 3 are tabulated below.

Table 3: Reconstructed optical constants of ruthenium ($\rho$=12.1 g/ cm$^3$), with their corresponding expanded uncertainty.

| λ/ nm | δ | β | λ/ nm | δ | β |
|---|---|---|---|---|---|
| 8 | 0.02615 ± 0.0006 | 0.004872 ± 0.000606 | 16 | 0.17638 ± 0.003886 | 0.066752 ± 0.004065 |
| 8.25 | 0.02816 ± 0.00064 | 0.004806 ± 0.000621 | 16.25 | 0.18232 ± 0.004019 | 0.075082 ± 0.004205 |
| 8.5 | 0.03031 ± 0.000685 | 0.004966 ± 0.000678 | 16.5 | 0.18771 ± 0.004153 | 0.084057 ± 0.004354 |
| 8.75 | 0.0325 ± 0.000737 | 0.004972 ± 0.000717 | 16.75 | 0.19241 ± 0.004278 | 0.093505 ± 0.004506 |
| 9 | 0.0348 ± 0.000781 | 0.005055 ± 0.000755 | 17 | 0.19646 ± 0.004391 | 0.103395 ± 0.004622 |
| 9.25 | 0.03731 ± 0.000838 | 0.005129 ± 0.000817 | 17.25 | 0.20008 ± 0.004541 | 0.113949 ± 0.004742 |
| 9.5 | 0.03987 ± 0.00089 | 0.005143 ± 0.000851 | 17.5 | 0.20282 ± 0.004646 | 0.124693 ± 0.004859 |
| 9.75 | 0.04265 ± 0.000949 | 0.005254 ± 0.00092 | 17.75 | 0.20443 ± 0.004731 | 0.135086 ± 0.004927 |
| 10 | 0.04551 ± 0.00101 | 0.005247 ± 0.000968 | 18 | 0.20532 ± 0.004837 | 0.146073 ± 0.005007 |
| 10.25 | 0.04852 ± 0.00107 | 0.005202 ± 0.001031 | 18.25 | 0.20527 ± 0.004877 | 0.157048 ± 0.00509 |
| 10.5 | 0.05178 ± 0.00114 | 0.005151 ± 0.001096 | 18.5 | 0.20396 ± 0.004945 | 0.167727 ± 0.005112 |
| 10.75 | 0.05532 ± 0.00121 | 0.005234 ± 0.001157 | 18.75 | 0.20187 ± 0.004994 | 0.177667 ± 0.005135 |
| 11 | 0.0591 ± 0.001296 | 0.005255 ± 0.00125 | 19 | 0.19916 ± 0.005007 | 0.186941 ± 0.005152 |
| 11.25 | 0.0632 ± 0.001373 | 0.005416 ± 0.001315 | 19.25 | 0.19585 ± 0.005033 | 0.19558 ± 0.005136 |
| 11.5 | 0.06737 ± 0.001468 | 0.006307 ± 0.001421 | 19.5 | 0.1926 ± 0.005033 | 0.203136 ± 0.005127 |
| 11.75 | 0.07167 ± 0.001553 | 0.006255 ± 0.001495 | 19.75 | 0.18964 ± 0.005071 | 0.209678 ± 0.005145 |
| 12 | 0.07658 ± 0.001666 | 0.006822 ± 0.001608 | 20 | 0.18663 ± 0.005093 | 0.216402 ± 0.005136 |
| 12.25 | 0.08168 ± 0.001758 | 0.007607 ± 0.001715 | 20.25 | 0.18433 ± 0.005115 | 0.22294 ± 0.005158 |
| 12.5 | 0.08699 ± 0.001878 | 0.008582 ± 0.001816 | 20.5 | 0.18215 ± 0.005112 | 0.229732 ± 0.00518 |
| 12.75 | 0.09255 ± 0.001989 | 0.00976 ± 0.001953 | 20.75 | 0.17945 ± 0.005146 | 0.236793 ± 0.005151 |
| 13 | 0.0985 ± 0.002118 | 0.01126 ± 0.002074 | 21 | 0.17672 ± 0.005195 | 0.243586 ± 0.005201 |
| 13.25 | 0.10471 ± 0.002254 | 0.013179 ± 0.002222 | 21.25 | 0.17292 ± 0.005222 | 0.250597 ± 0.005199 |
| 13.5 | 0.11112 ± 0.002395 | 0.015511 ± 0.002375 | 21.5 | 0.16884 ± 0.005262 | 0.25795 ± 0.005166 |
| 13.75 | 0.11782 ± 0.002544 | 0.018358 ± 0.002541 | 21.75 | 0.1642 ± 0.005282 | 0.265002 ± 0.00514 |
| 14 | 0.12464 ± 0.002694 | 0.021672 ± 0.002705 | 22 | 0.1587 ± 0.005273 | 0.271337 ± 0.005097 |
| 14.25 | 0.13143 ± 0.002845 | 0.025593 ± 0.002881 | 22.25 | 0.15343 ± 0.005267 | 0.276512 ± 0.005055 |
| 14.5 | 0.13826 ± 0.003005 | 0.030114 ± 0.003059 | 22.5 | 0.14676 ± 0.005325 | 0.281401 ± 0.005003 |
| 14.75 | 0.14479 ± 0.003151 | 0.035135 ± 0.00322 | 22.75 | 0.1404 ± 0.005302 | 0.284277 ± 0.004895 |
| 15 | 0.15122 ± 0.003298 | 0.040448 ± 0.003398 | 23 | 0.13452 ± 0.005281 | 0.287145 ± 0.004843 |
| 15.25 | 0.15757 ± 0.003449 | 0.046254 ± 0.003563 | 23.25 | 0.13042 ± 0.005368 | 0.289225 ± 0.004836 |
| 15.5 | 0.16393 ± 0.003597 | 0.052531 ± 0.003739 | 23.5 | 0.12672 ± 0.00549 | 0.291547 ± 0.004854 |
| 15.75 | 0.1702 ± 0.003746 | 0.059297 ± 0.003896 | 23.75 | 0.12374 ± 0.005588 | 0.293875 ± 0.004897 |